\PassOptionsToPackage{dvipsnames,usenames}{xcolor} 

\documentclass[journal=jpclcd,manuscript=letter,articletitle=false,layout=twocolumn]{achemso}

\usepackage[colorlinks,pdfencoding=auto,psdextra]{hyperref}
\usepackage{doi} 

\usepackage{braket}
\usepackage[T1]{fontenc} 
\usepackage{amsmath} 
\usepackage{amssymb} 
\usepackage{placeins}
\usepackage[usenames, dvipsnames]{xcolor} 

\setkeys{acs}{doi = true}
\newcommand{\onlinecite}[1]{\hspace{-1 ex} \nocite{#1}\citenum{#1}} 

\author{E. Kirstein}
\affiliation{Experimentelle Physik 2, Department of Physics, TU Dortmund,  44227 Dortmund, Germany}
\email{erik.kirstein@tu-dortmund.de}

\author{D. R. Yakovlev}
\affiliation{Experimentelle Physik 2, Department of Physics, TU Dortmund, 44227 Dortmund, Germany} 
\alsoaffiliation{Ioffe Institute, Russian Academy of Sciences, 194021 St. Petersburg, Russia}
\email{dmitri.yakovlev@tu-dortmund.de}

\author{E. A. Zhukov}
\affiliation{Experimentelle Physik 2, Department of Physics, TU Dortmund, 44227 Dortmund, Germany}
\alsoaffiliation{Ioffe Institute, Russian Academy of Sciences, 194021 St. Petersburg, Russia}

\author{J. H\"ocker}
\affiliation{Experimental Physics VI, Julius-Maximilian University of W\"urzburg, 
97074 W\"{u}rzburg, Germany}

\author{V. Dyakonov}
\affiliation{Experimental Physics VI, Julius-Maximilian University of W\"urzburg, 
97074 W\"{u}rzburg, Germany}

\author{M. Bayer}
\affiliation{Experimentelle Physik 2, Department of Physics, TU Dortmund, 44227 Dortmund, Germany} 
\alsoaffiliation{Ioffe Institute, Russian Academy of Sciences, 194021 St. Petersburg, Russia}

\abbreviations{IR,NMR,UV}

\title[MAPbI Spin Dynamics] {Spin dynamics of electrons and holes interacting with nuclei in MAPbI$_3$ perovskite single crystals}

\begin{document}

\makeatletter
\setlength\acs@tocentry@height{0.8\columnwidth}
\setlength\acs@tocentry@width{0.8\columnwidth}
\makeatother

\begin{tocentry}

\includegraphics[width=\columnwidth]{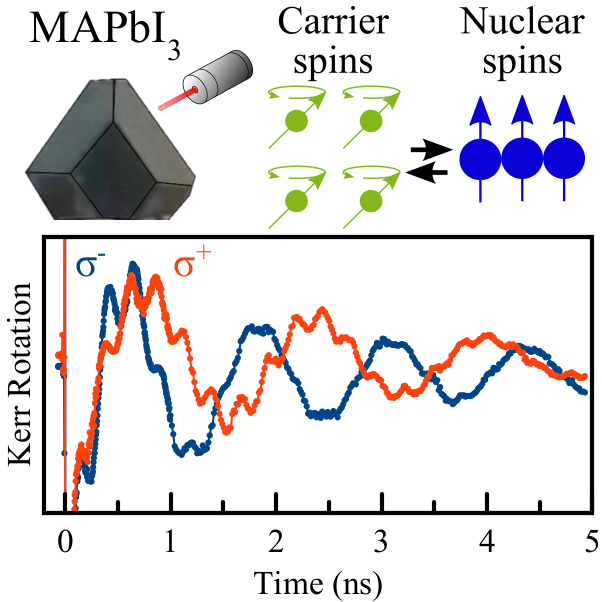}

\end{tocentry}

\begin{abstract}
Methylammonium lead triiodine (MAPbI$_3$) is a material representative of the hybrid organic-inorganic lead halide perovskites which attract currently great attention due to their photovoltaic efficiency and bright optoelectronic properties. Here, the coherent spin dynamics of charge carriers and spin dependent phenomena induced by the carrier interaction with nuclear spins are studied in MAPbI$_3$ single crystals, using time-resolved Kerr rotation at cryogenic temperatures in magnetic fields up to 3~T. Spin dephasing times up to a few nanoseconds and a longitudinal spin relaxation time of 37~ns are measured. The Larmor spin precession of both resident electrons and holes is identified in the Kerr rotation signals. The Land\'e factors ($g$-factors) in the orthorhombic crystal phase show a strong anisotropy, ranging for the holes from $-0.28$ to $-0.71$ and for the electrons from $+2.46$ to $+2.98$, while the $g$-factor dispersion of about 1\% is rather small. An exciton $g$-factor of $+2.3$ is measured by magneto-reflectivity. A dynamic nuclear polarization by means of spin polarized electrons and holes is achieved in tilted magnetic fields giving access to the carrier-nuclei exchange interaction and the nuclei spin relaxation time exceeding 16 minutes. 
\end{abstract}


Since the proposal to use hybrid organic-inorganic lead halide perovskites for fabricating efficient solar cells in 2009 \cite{kojima2009}, the research activities on these materials have rapidly increased, resulting in remarkable achievements. A photovoltaic efficiency exceeding 25\% is reached~\cite{NREL2021,Qiu2021}, also long-standing challenges, like the limited lifetime of perovskite solar cells, come closer to a solution~\cite{zhao2021}. Further, these materials and their nanostructures look promising for application in optoelectronic devices~\cite{fu2019,jena2019}, X-ray detectors \cite{nazarenko2017,wei2019}, and spintronics devices~\cite{wang2019,ning2020,zhang2015,kim2021}. These perspectives demand a solid understanding of the fundamental properties of this class of perovskites, for which optical and magneto-optical techniques developed for the investigation of spin-dependent phenomena are valuable tools, as they give access to key band structure parameters~\cite{yu2016,baranowski2020,piveteau2020,kirstein2021}. 

Among the experimental techniques used for studying the spin physics of semiconductors, time-resolved pump-probe Faraday (Kerr) rotation is one of those providing most information~\cite{yakovlevCh6,glazov2018}. It allows one to measure the Land\'e factors ($g$-factors), study spin relaxation, assess spin coherence and spin dephasing times of electrons and holes and to assess their interaction with the nuclear spin system. Applying additionally optically detected nuclear magnetic resonance (ODNMR), the identification of the contained nuclear isotopes interacting with the carriers is possible~\cite{zhukov2014,heisterkamp2015b,kirstein2021}. Time-resolved Kerr rotation (TRKR) has been used to study the perovskite single crystals CsPbBr$_3$~\cite{belykh2019} and FA$_{0.9}$Cs$_{0.1}$PbI$_{2.8}$Br$_{0.2}$~\cite{kirstein2021}, the polycrystalline films MAPbI$_{3-x}$Cl$_x$~\cite{odenthal2017}, MAPbI$_3$~\cite{garcia-arellano2021} and CsPbBr$_3$~\cite{grigoryev2021}, and the nanocrystals CsPbBr$_3$~\cite{crane2020,grigoryev2021}.

In this paper we apply TRKR to study the coherent spin dynamics of electrons and holes and their interaction with the nuclear spin system in MAPbI$_3$ (CH$_3$NH$_3$PbI$_3$) single crystals. We measure the electron and hole $g$-factors, their anisotropy and dispersion, and their spin dephasing and longitudinal spin relaxation times. The exciton $g$-factor is measured by magneto-reflectivity. Dynamic nuclear polarization of $^{207}$Pb nuclear isotope is achieved and a minutes-long spin relaxation time of the nuclei is measured. The spin dynamics parameters are compared with the available data for a MAPbI$_3$ polycrystalline film and for bulk single perovskite crystals of different composition.

The paper is organized as follows. We start with the results on the magneto-reflectivity and photoluminescence of MAPbI$_3$ focusing on its exciton parameters. Then we turn to TRKR data to study the coherent spin dynamics of electrons and holes and their anisotropic $g$-factors. Finally we present results on the carrier interaction with nuclear spins and on the spin dynamics of the nuclear spin system. 

\section{Exciton spectroscopy}

\begin{figure*}[!ht]
  \includegraphics[width=\textwidth]{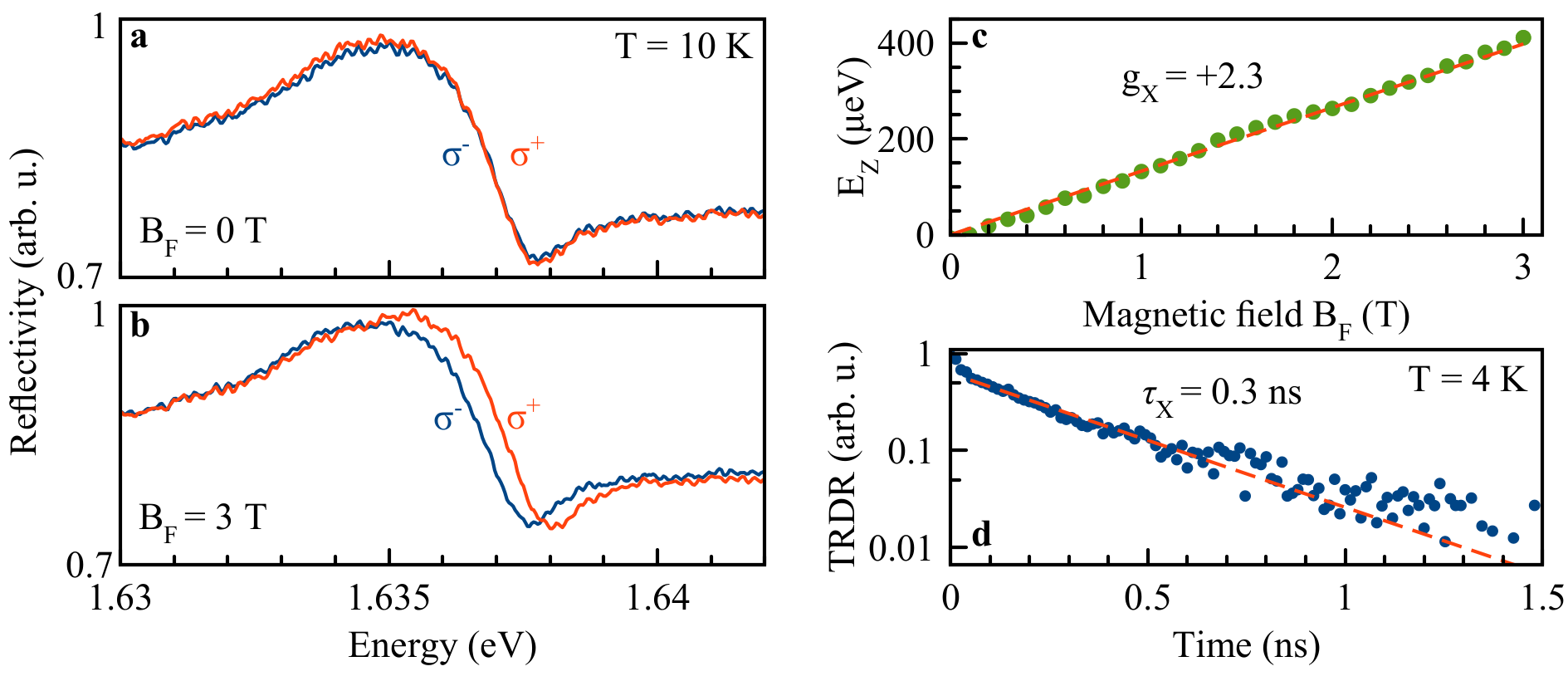}
  \caption{Reflectivity of the MAPbI$_3$ single crystal. (a) Reflectivity spectra of the exciton at zero magnetic field in $\sigma^+$ (red) and $\sigma^-$ (blue) circular polarizations. (b) Reflectivity spectra in the magnetic field of 3~T applied in the Faraday geometry. (c) Magnetic field dependence of the exciton Zeeman splitting evaluated from the reflectivity spectra. Symbols are experimental data and line is linear fit corresponding to the exciton $g$-factor $g_\textrm{X}=+2.3$. (d) Time-resolved differential reflectivity (TRDR) measured at the energy of 1.638~eV. Dashed line is single exponential fit with $\tau_\textrm{X}=0.3$~ns. $B=0$~T.}
  \label{fig:exgfactor}
\end{figure*}

The studied MAPbI$_3$ crystal shows a pronounced exciton resonance in the reflectivity spectrum measured at $T=10$~K~\cite{hopfield1963}, see Figure~\ref{fig:exgfactor}a. The resonance energy is 1.6365~eV and the exciton damping is 3~meV. In magnetic field applied in the Faraday geometry, the exciton resonance has a Zeeman splitting, $E_\textrm{Z}$, as one can see from the $\sigma^+$ and $\sigma^-$ circularly polarized spectra in Figure~\ref{fig:exgfactor}b. The Zeeman splitting increases linearly with magnetic field, allowing us to define the value and sign of the exciton $g$-factor, $g_\textrm{X}=+2.3$. This is shown in Figure~\ref{fig:exgfactor}c, where the linear fit corresponds to $E_\textrm{Z}=g_\textrm{X}\mu_\textrm{B}B_{\rm F}$. Here $B_{\rm F}$ is magnetic field strength and $\mu_\textrm{B}$ the Bohr magneton. This exciton $g$-factor is in agreement with $g_\textrm{X}=2.66$, measured in transmission of MAPbI$_3$ crystals in strong magnetic fields up to 66~T~\cite{yang2017}.   

The exciton population dynamics was measured at $T=4$~K by means of time-resolved differential reflectivity (TRDR). The laser photon energy of 1.638~eV was tuned to the exciton resonance. The TRDR signal shows to a good approximation a monoexponential decay with time $\tau_\textrm{X}=0.3$~ns, see Figure~\ref{fig:exgfactor}d, which can be related to the exciton lifetime. 

\begin{figure}[!ht]
  \includegraphics[width=.99\columnwidth]{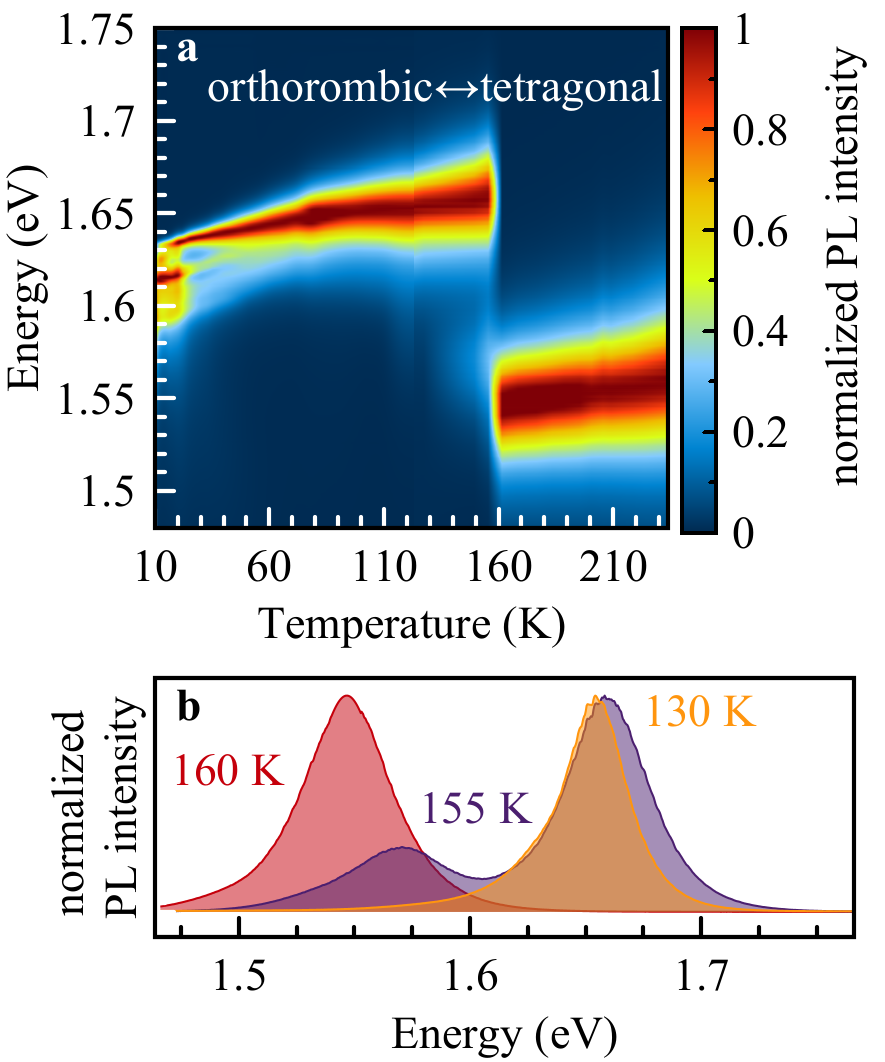}
  \caption{(a) Two-dimensional plot of the temperature evolution of photoluminescence spectra taken from the MAPbI$_3$ crystal. Excitation energy is 3.06~eV (405~nm). (b) PL spectra at $T=130$, 155 and 160~K in vicinity of the structural phase transition. }
  \label{fig:phasetrans}
\end{figure}

It is important for our study, that at temperatures below 160~K MAPbI$_3$ crystals exhibit an orthorhombic crystal structure~\cite{whitfield2016}, which causes an anisotropy of the spin dependent properties, e.g., the $g$-factor anisotropy for electrons and holes shown below. The structural phase transition with decreasing temperature from the tetragonal to the orthorhombic phase is accompanied by a step-like increase of the band gap by 113~meV undercutting the temperature of about 160~K~\cite{whitfield2016}. In optical experiments it is evidenced by a corresponding high energy shift of the photoluminescence (PL) line, which is shown in Figure~\ref{fig:phasetrans}a. Here the evolution of PL spectra in the temperature range from 10 to 235~K is presented as two-dimensional contour plot, where the PL intensity is encoded by colors. One can see that the band gap decreases with lowering temperature, as typical for lead halide perovskites~\cite{lopez2020}, except of the high energy shift at the phase transition. Some PL spectra above and below the phase transition are also shown in Figure~\ref{fig:phasetrans}b.  

\section{Coherent spin dynamics of electrons and holes}

\begin{figure*} [ht!]
 \includegraphics[width=\textwidth]{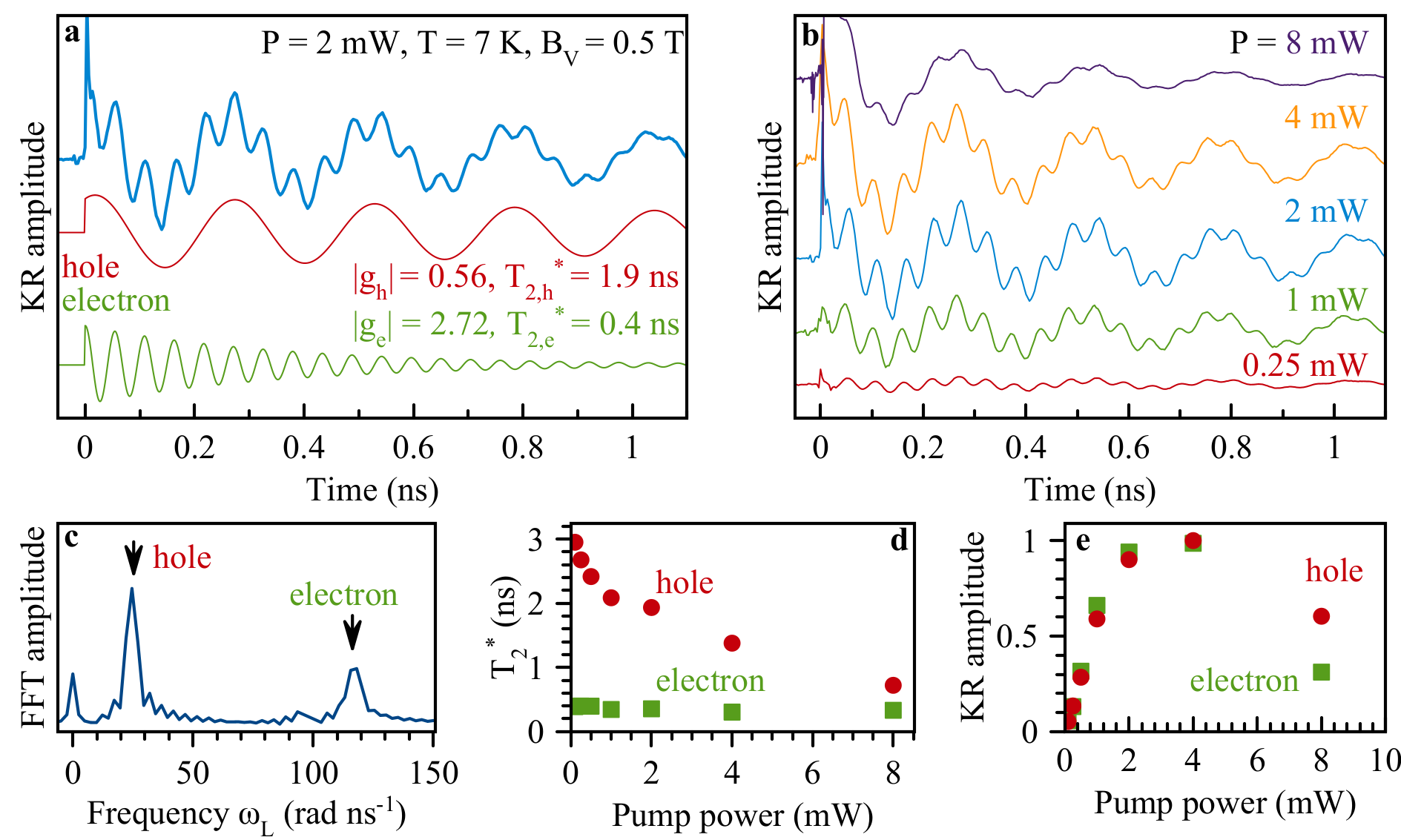}
 \caption{Coherent spin dynamics of electrons and holes in the MAPbI$_3$ crystal at $T=7$~K. (a) TRKR signal at $B_\textrm{V}=0.5$~T (blue) contributed by electron (red) and hole (green) oscillating components shown separated below. Pump power is 2~mW, laser photon energy is $1.6359$~eV. 
(b) TRKR signals for various pump powers. The signals are shifted vertically for clarity, but not normalized in amplitude. 
(c) Fast Fourier transformation (FFT) spectrum of the TRKR signal from panel (a). The two distinct peaks attributed to electron ($\omega_{{\rm L},e}=119.4$~rad ns$^{-1}$) and hole ($\omega_{{\rm L},h}=24.6$~rad ns$^{-1}$) spin precession are indicated by arrows. 
(d) Pump power dependence of the electron and hole spin dephasing times.
(e) KR amplitude of the electron (green, $S_e$) and hole (red, $S_h$) components at $t=0$ as a function of pump power.}
 \label{fig:Voigt}
\end{figure*}

To study the electron and hole spin coherence, we implement the TRKR technique. In this experiment, carrier spin polarization is generated optically by circularly polarized pump pulses. Its Larmor precession about the magnetic field $\mathbf{B}_{\rm V}$ applied in the Voigt geometry is detected via the Kerr rotation (KR) of linear polarization of the probe pulses~\cite{yakovlevCh6}. A typical TRKR signal of the studied MAPbI$_3$ crystal, measured at $T=7$~K and $B_\textrm{V}=0.5$~T, is shown in Figure~\ref{fig:Voigt}a. One can see that the signal contains two oscillating components with quite different frequencies so that they can be distinguished in the fast Fourier transformation (FFT) spectrum (Figure~\ref{fig:Voigt}c). These components are also shown separately in Figure~\ref{fig:Voigt}a. Using an appropriate fit function for the KR amplitude, $A_{\rm KR}(t)=\sum_{e,h}{S_{e(h)}} {\cos (\omega_{{\rm L},e(h)} t)} \exp(-t/T^*_{2,e(h)})$, with two damped cosine functions (see Methods) we evaluate the electron and hole parameters. The faster precession can be assigned to the electron spins with the $g$-factor $g_{e}=\hbar\omega_{{\rm L},e}/ \mu_B B_{\rm V}=+2.72$ and the dephasing time $T_{2,e}^*=0.4$~ns, and the slower one to the hole spins with $g_{h}=-0.56$ and $T_{2,h}^*=1.9$~ns. This assignment is based on the values of the $g$-factors, as for MAPbI$_3$ it is known that $|g_e| \gg |g_h|$, for details see Ref.~\onlinecite{kirstein2021nc}. TRKR provides information on the $g$-factor value, but not on its sign. Information on the sign is obtained from model calculations predicting for MAPbI$_3$ $g_e>0$~\cite{yu2016,kirstein2021nc}. The knowledge about the exciton $g$-factor including its sign ($g_{\rm X}=+2.3$) allows us to identify the negative sign of $g_h$ using $g_{\rm X}=g_e+g_h$~\cite{kirstein2021nc}. The opposite signs of the electron and hole $g$-factors in MAPbI$_3$ are confirmed by our results on dynamic nuclear polarization presented below.

With regard to their dependencies on temperature, magnetic field, and pump power, which we will present below, the TRKR signals in the studied MAPbI$_3$ sample in general are similar to the behavior reported for other lead halide perovskites, like CsPbBr$_3$~\cite{belykh2019}, FA$_{0.9}$Cs$_{0.1}$PbI$_{2.8}$Br$_{0.2}$~\cite{kirstein2021}, MAPbI$_{3-x}$Cl$_x$~\cite{odenthal2017}, and MAPbI$_3$~\cite{garcia-arellano2021}. The intensely debated problem on whether the TRKR signals are provided by electrons and holes within the excitons or originate from long-living electrons and holes localized in different crystal sites was resolved by strong experimental facts in favor of the carrier localization~\cite{belykh2019,kirstein2021,garcia-arellano2021}. Among them are: (i) the long spin dephasing times $T_2^*$ of the KR signals exceeding the exciton lifetime, (ii) the strong temperature dependence of $T_2^*$, (iii) the linear dependencies of the Larmor frequencies on the magnetic field down to very weak magnetic fields below 10~mT (note a weak regular periodic signal is observable in the resonant spin amplification regime~\cite{yakovlevCh6}, but not shown here), confirming the absence of electron-hole exchange interaction effects, (iv) the spatial and spectral variation of the relative intensities of the electron and hole signals. Those features were confirmed as well for the studied MAPbI$_3$ crystal. Therefore, we attribute the TRKR signal to resident electrons and holes, which can be provided by photogeneration followed by their spatial separation and localization. In this case, the mechanism of generating spin coherence of resident carriers is the same as previously considered for III-V and II-VI semiconductor quantum wells hosting a two-dimensional electron gas of low density or for singly-charged (In,Ga)As/GaAs quantum dots~\cite{yakovlevCh6}.

TRKR signals measured for various pump powers in the range $P=0.25-8$~mW are shown in Figure~\ref{fig:Voigt}b. The spin dephasing times shorten for increasing power, for the holes from 3~ns at 0.1~mW down to 0.8~ns at 8~mW and for the electrons from 0.5~ns down to 0.2~ns across the same power range, Figure~\ref{fig:Voigt}d. The electron and hole signal amplitudes are about equal ($S_e \simeq S_h$) in the whole power range. But their values show a pronounced nonmonotonic dependence, see Figure~\ref{fig:Voigt}e. They increase strongly with power in the low power regime up to 2~mW, then saturate and subsequently strongly drop at the highest used pump power $P=8$~mW. Such power dependencies of the carrier spin dephasing times and their KR amplitudes can be explained by heating and carrier delocalization. Note, the Larmor frequencies are independent of the pump power.

We also measured the dependencies of the KR amplitudes and spin dephasing times on the bath (lattice) temperature for electrons and holes at $B_V=0.5$~T, but do not show them here. The signal amplitudes decrease strongly for temperature increasing up to $T=25$~K. The spin dephasing times shorten as well with growing temperature, down to 55~ps at $T=25$~K, following an Arrhenius-like activation dependence, $T_{2}^*(T)^{-1}= 1/T_0 + w \exp{(-E_A/k_{\rm B} T)}$, with the activation energy $E_A$, the thermal relaxation rate $w$, and the temperature independent dephasing time $T_0$. The activation energies are in the range from 1.4 to 4.3~meV.

\begin{figure} [ht!]
  \includegraphics[width=.99\columnwidth]{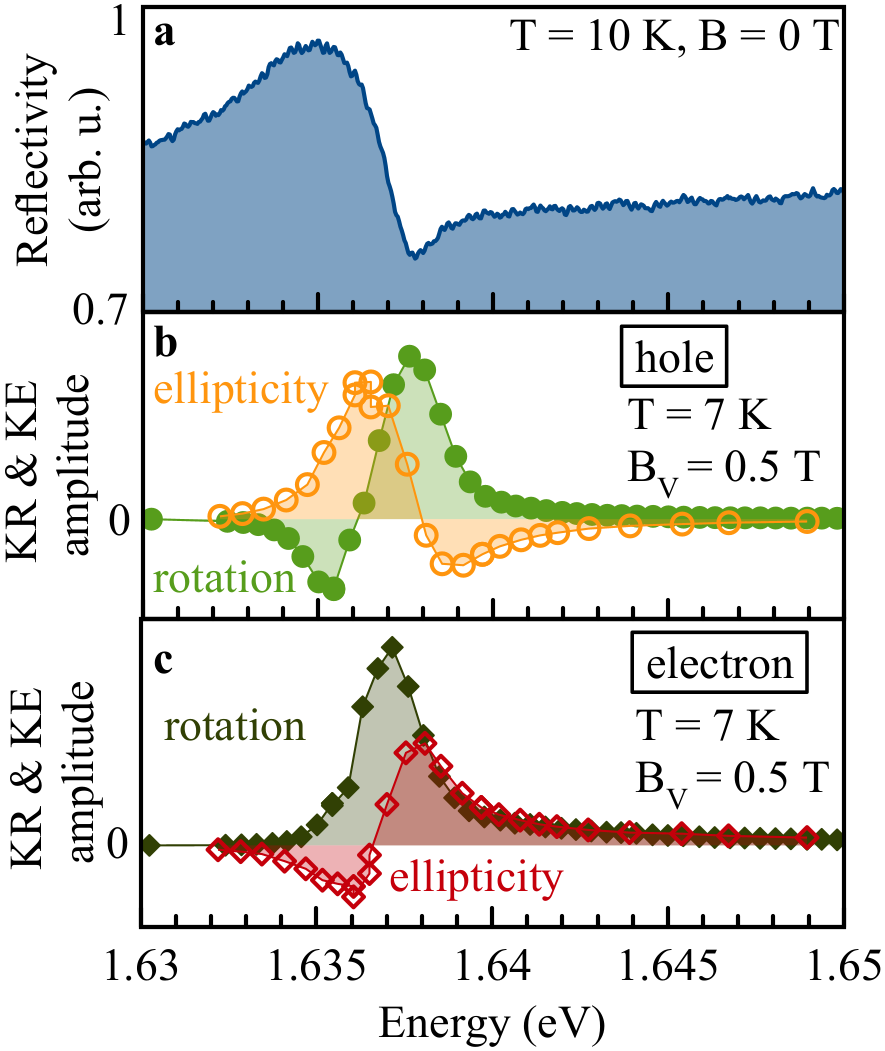}
  \caption{(a) Reflectivity spectrum of the MAPbI$_3$ crystal given as a reference (same as in Figure~\ref{fig:exgfactor}a). (b) Spectral dependencies of Kerr rotation (green) and Kerr ellipticity (yellow) amplitudes of the hole signals. (c) Spectral dependencies of Kerr rotation (dark green) and Kerr ellipticity (dark red) amplitudes of the electron signals. The results in panels (b,c) were measured at $B_\textrm{V}=0.5$~T for $T=7$~K. The pump and probe powers are 0.25~mW and 0.1~mW, respectively.}
\label{fig:KerrSpec}
\end{figure}

In Figure~\ref{fig:KerrSpec}, the spectral dependencies of the KR and Kerr ellipticity (KE) signals are shown for electrons and holes. The reflectivity spectrum in Figure~\ref{fig:KerrSpec}a is given for comparison. One can see that the spin coherent signals are observed only in the vicinity of the exciton resonance, demonstrating that the exciton plays an important role in the generation of the carrier spin coherence and also in its detection. The typical full width at half maximum (FWHM) of the KR and KE signals are about 2.5~meV, similar to the width of the exciton resonance in reflectivity. Note that the width of the exciton resonance in the studied single crystal is by about an order of magnitude smaller than the one of 20~meV in a MAPbI$_3$ polycrystalline film~\cite{garcia-arellano2021}. 

The model considerations in Refs.~\onlinecite{yugova2009,glazov2018} show that in case of a single resonance, the KR spectral dependence has the shape of the derivative of the absorption resonance with zero amplitude at the resonance energy, whereas the KE follows the absorption resonance. One can see in Figures~\ref{fig:KerrSpec}b,c that the measured signals deviate from these model predictions, both for the electrons and the holes. This evidences that a dedicated analysis is necessary to describe the KR and KE spectral profiles. This analysis should take into account the localization potentials of the resident carriers and possible contributions of bound exciton states, which goes beyond the scope of this paper.

\section{\texorpdfstring{$g$}{g}-factor anisotropy}

\begin{figure*} [ht!]
  \includegraphics[width=\textwidth]{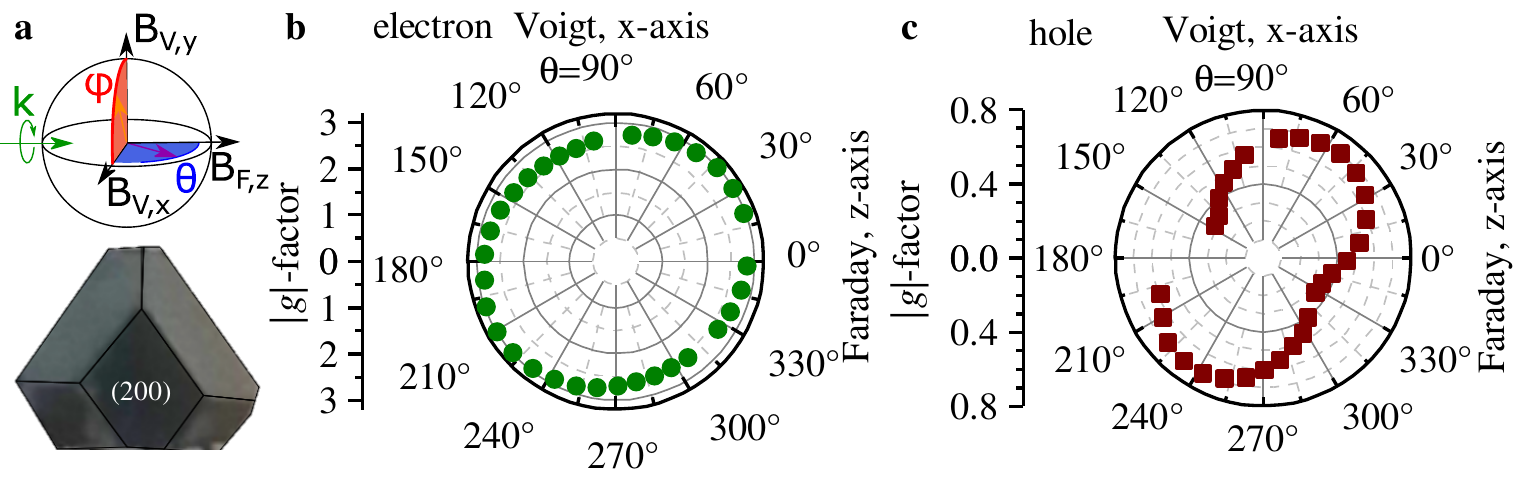}
  \caption{$g$-factor anisotropy in the MAPbI$_3$ crystal. (a) Sketch of the measurement geometry, where $\textbf{k} \parallel z$ is the light wave vector, $\mathbf{B}_{\rm F} \parallel \textbf{k}$ is the magnetic field in the Faraday geometry and $\mathbf{B}_{\rm V} \perp \textbf{k}$ is the magnetic field in the Voigt geometry. Below a sketch of the sample with edge highlighted lines is shown. The front facet is the (200) crystallographic plane ($[200]\parallel \textbf{k}$). (b-c) $g$-factor anisotropies of the electrons and holes measured by magnetic field rotation in the Faraday-Voigt plane by tuning the angle $\theta$, while keeping the angle $\varphi=0^\circ$ constant. The magnetic field strength is kept constant at $B=0.5$~T. $T=7$~K.}
  \label{fig:aniso}
\end{figure*}

TRKR experiments performed in a vector magnet allow us to address the anisotropy of the electron and hole $g$-factors, provided by the structural anisotropy. In our experiment, the MAPbI$_3$ crystal is oriented such that the $a$-axis (\{200\}), in tetragonal (room temperature) crystallographic configuration, is parallel to the light wave vector $\textbf{k}$, see Figure~\ref{fig:aniso}a. The magnetic field with a fixed strength of $B=0.5$~T is rotated in the Faraday-Voigt plane (horizontal plane) by tuning angle $\theta$ for fixed $\varphi=0^\circ$.

Rotational diagrams of the electron and hole $g$-factors are shown in Figures~\ref{fig:aniso}b,c. Both $g$-factors are anisotropic: $g_e$ ranges from $+2.56$ ($\theta=120^\circ$) to $+2.98$ ($30^\circ$) and $g_h$ from $-0.33$ ($150^\circ$) to $-0.68$ ($60^\circ$). Surprisingly, for both electron and hole, the $g$-factor in-plane symmetry axes do not coincide with the crystallographic $a$-axis $\theta=0^\circ$. The reason for that is not clear yet, some tentative explanations can be found in Ref.~\onlinecite{kirstein2021nc}. A further tilting of the magnetic field out of the Faraday-Voigt plane by tuning the angle $\varphi$ allows us to find the orientation of the main long axis of the $g_e$ tensor at $\theta=33^\circ$, $\varphi=54^\circ$ with extremal values ranging from $+2.46$ to $+2.98$. For the $g_h$ tensor it has the orientation $\theta=57^\circ$, $\varphi=54^\circ$ and values in the range from $-0.28$ to $-0.71$. In relative numbers, the $g$-factor anisotropy is strong for the holes amounting to 47\% compared to 10\% for the electrons. Note, the $g$-factor tensor, in terms $g_{x'}\approx g_{y'} \neq g_{z'}$, is in general good agreement with MAPbI$_3$ low temperature orthorhombic Pnma configuration $a \approx b \neq c$-axis \cite{whitfield2016}, with $\{x',y',z'\}$-prime axis being a tilted coordinate system, rotated to $g$-factor extremal values.

\begin{figure*}[ht!]
  \includegraphics[width=\textwidth]{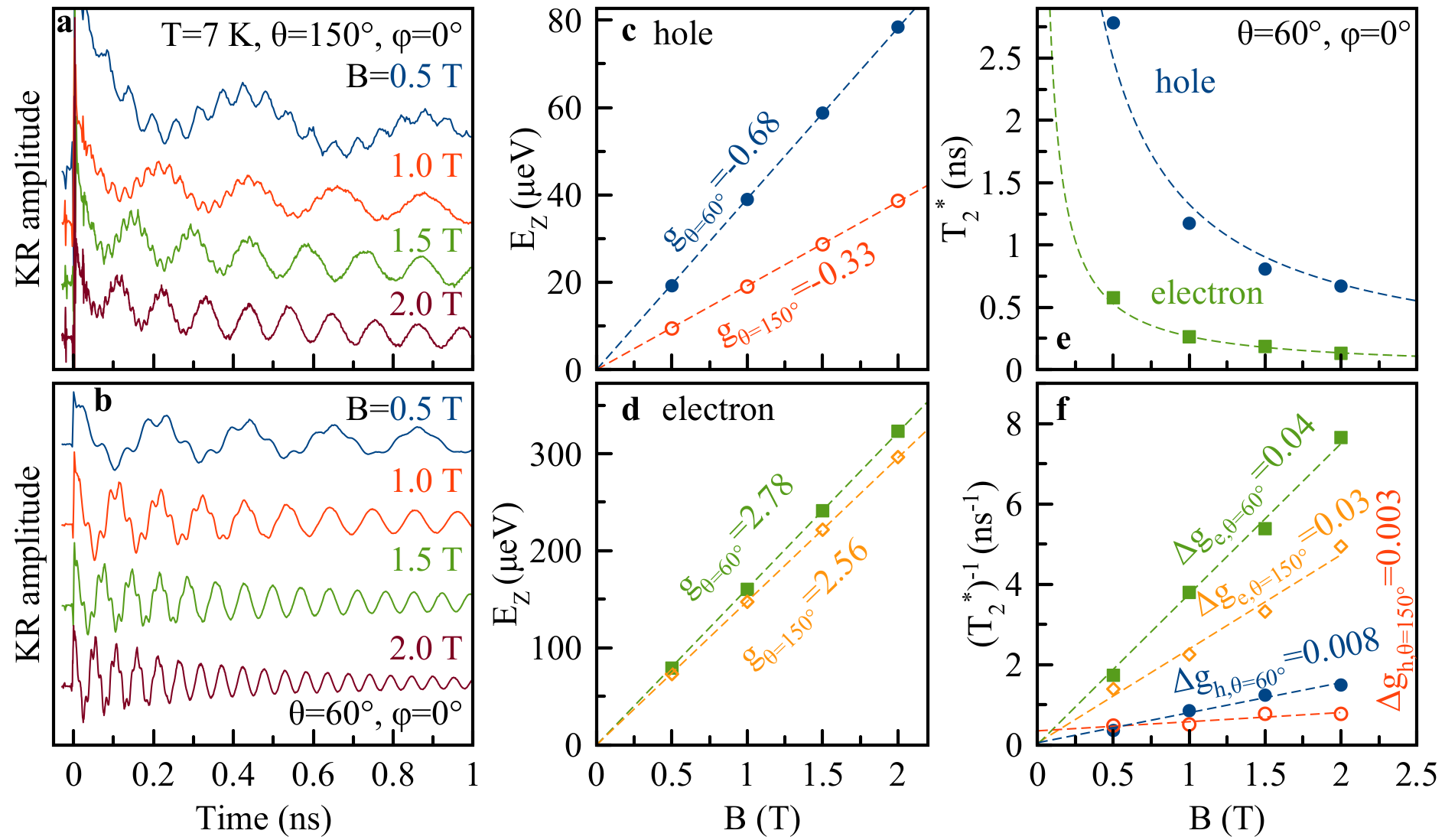}
  \caption{Magnetic field dependence of the coherent spin dynamics. (a,b) TRKR signals at $\theta=150^\circ$ and $60^\circ$, with each $\varphi=0^\circ$, for increasing magnetic field strength. The spin parameters obtained from fitting the TRKR curves are plotted in the others panels, namely by blue and red [green and yellow] circles for holes [electrons] at $\theta=60^\circ$ and $150^\circ$, respectively. (c,d) Magnetic field dependencies of the Zeeman splitting ($E_\textrm{Z}$) for holes and electrons. Special attention has to be paid to the absence of an offset in the linear fit at zero field (dashed lines). (e) Magnetic field dependence of the spin dephasing time $T_2^*$, data (symbols) and fits (dashed lines), at $\theta=60^\circ$. (f) Spin dephasing rate as function of the magnetic field for the two field orientations. Linear dependencies (dashed lines) were fitted to the data (symbols) to extract the dispersion of the $g$-factors $\Delta g$, with corresponding color coded labels in the panel. $T=7$~K.}
  \label{fig:Bdep}
\end{figure*}

The evolution of the TRKR signals with rising magnetic field strength is illustrated in Figures~\ref{fig:Bdep}a,b. For that, we chose two field orientations, namely $\theta=60^\circ$ and $150^\circ$, where for $\varphi=0^\circ$ the hole $g$-factor has its extremal values. For $\theta=150^\circ$, the KR signal is superimposed on a nonoscillating exponential component, provided by a spin polarization with a projection of the $z$-axis, i.e. parallel to $\mathbf{k}$. One can see that the Larmor precession frequencies, $\omega_{{\rm L},e(h)}$, increase with growing magnetic field, reflecting the increase of the carrier Zeeman splittings plotted in Figures~\ref{fig:Bdep}c,d. It is important to note here that the field dependencies of the splittings extrapolated to zero magnetic field go to zero. This is in line with our assignment of the KR signals to resident electrons and holes not bound in excitons. In the latter case, a finite splitting provided by the electron-hole exchange interaction would be expected.

The spin dephasing times, $T_{2,e(h)}^*$, shorten with increasing magnetic field. Their field dependence is given in Figure~\ref{fig:Bdep}e. These times are inversely proportional to the magnetic field, which is better seen when the magnetic field dependence of the spin dephasing rates are plotted in Figure~\ref{fig:Bdep}f. We fit these data using the equation
\begin{equation}
1/T_{2,e(h)}^*(B) = 1/T_{2,0,e(h)}^* + \Delta g_{e(h)} \mu_{\rm B} B / \hbar.
\end{equation}
Here $\Delta g_{e(h)}$ are the dispersions of the $g$-factors and $T_{2,0,e(h)}^*$ are the spin dephasing times independent of the external magnetic field~\cite{yakovlevCh6}. The fit allows us to evaluate the $g$-factor dispersions which show also strong anisotropies. For the holes $\Delta g_h/g_h (150^\circ) = 0.003/0.33=0.9$\% and $\Delta g_h/g_h (60^\circ) = 0.008/0.68=1.2$\%, while for the electrons $\Delta g_e/g_e(150^\circ) = 0.03/2.56=1.2$\% and $\Delta g_e/g_e (60^\circ) = 0.04/2.78=1.4$\%. As one can see from the given values, the $\Delta g$ anisotropy follows the anisotropy of the $g$ values, limiting the relative anisotropy values $\Delta g/g$ to about 1\%.

\section{Dynamic nuclear polarization}

\begin{figure*}[ht!]
  \includegraphics[width=\textwidth]{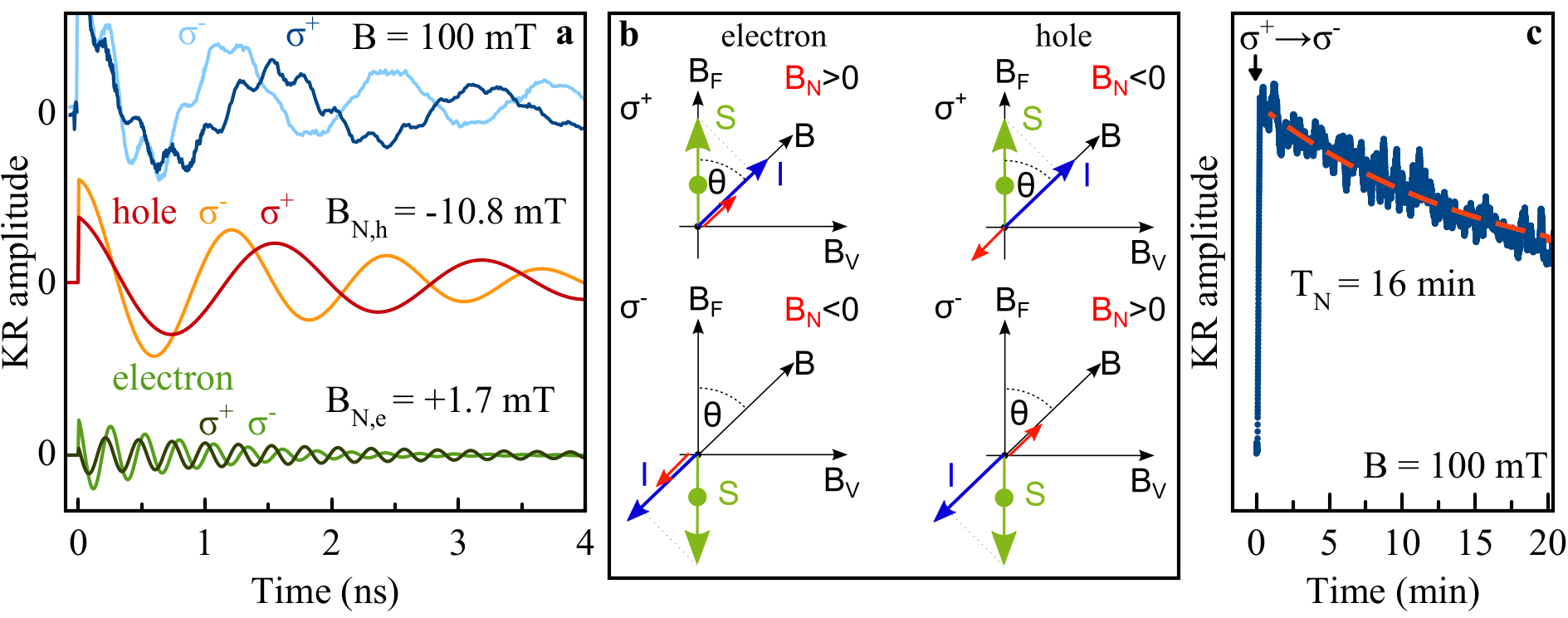}
  \caption{Dynamic nuclear polarization. (a) TRKR signals for constant $\sigma^+$ (dark blue) or $\sigma^-$ (light blue) pump helicity, measured at $B=100$~mT tilted to $\theta=60^\circ$ ($\varphi=0^\circ$). $T=1.6$~K, 15~mW pump power. The signals decomposed into the hole and electron contributions are shown below. They clearly differ in the Larmor precession frequencies for the two opposite circular polarizations due to the Overhauser field of polarized nuclei.  
(b) Sketch of the build up of dynamic nuclear polarization in MAPbI$_3$ according to eq.~\ref{eq:DNP}. For a $\sigma^+$ polarized pump, the nuclear spin polarization $\mathbf{I}$ (blue arrow) manifests itself as a Overhauser field $\mathbf{B}_N$ (red arrow), which for electrons ($g_e>0$) is directed along $\mathbf{I}$ and $\mathbf{B}$, while for holes ($g_h<0$) $\mathbf{B}_N$ points along the opposite direction. 
(c) Dynamics of the DNP after changing the pump helicity from $\sigma^+$ to $\sigma^-$ measured at the time delay of 3.3~ns. The other experimental conditions are the same as in panel (a). Dashed red line is a monoexponential fit, giving a nuclear spin polarization time of $T_{N}=16$~min.}
  \label{fig:DNP}
\end{figure*}

For charge carrier spins in semiconductors, the hyperfine interaction with the nuclear spins plays an important role in spin polarization and spin dynamics~\cite{glazov2018}. This is also valid for the lead halide perovskites, as demonstrated recently for FA$_{0.9}$Cs$_{0.1}$PbI$_{2.8}$Br$_{0.2}$ crystals~\cite{kirstein2021}. In this manuscript, a theoretical model of the carrier-nuclei interaction in the perovskites was developed, where the dominant role of the $^{207}$Pb isotope in this interaction was established theoretically and experimentally. We base the present study on these results, as MAPbI$_3$ has strong similarity to FA$_{0.9}$Cs$_{0.1}$PbI$_{2.8}$Br$_{0.2}$ in this respect.  

In the lead halide perovskites, e.g. MAPbI$_3$, the electronic states at the top of the valence band and at the bottom of the conduction band are formed by atomic orbitals of lead and iodine, respectively, without notable contributions from the MA-cation~\cite{kirstein2021,nestoklon2021}. Their nuclei have isotopes with nonzero spin: iodine $^{127}$I has 100\% abundance of spin 5/2 and lead $^{207}$Pb 22.2\% abundance with spin 1/2~\cite{piveteau2020}. The valence band Bloch wavefunctions have a considerable contribution of the $s$-type lead orbitals, which dominate the hyperfine coupling with the $^{207}$Pb isotope characterized by the hyperfine coupling constant of about $A=100$~$\mu$eV ($A/h=242$~GHz). The conduction band Bloch wavefunctions are mainly contributed by the $p$-type lead orbitals with some contribution of $s$-type iodine orbitals. Their effect results in the electron hyperfine coupling constant to be on the order of several $\mu$eV.

In order to study the carrier-nuclei interaction we modified the experiment to obtain dynamic nuclear polarization (DNP). In the studies so far, the pump polarization helicity was modulated so fast that no nuclear polarization could build up. In the present experiment the pump helicity is fixed. Then the spin polarization of the carriers, which are optically oriented by the pump beam, is transferred to the nuclear spin system (NSS), which becomes polarized. The nuclear spin polarization $\braket{\textbf{I}}$ is described by~\cite{meier_optical_1984} 
\begin{equation}
\label{eq:DNP}
		\braket{\textbf{I}} = l \frac{4I(I+1)}{3} \frac{\textbf{B}(\textbf{B}\cdot\braket{\textbf{S}_{e(h)}})}{B^2} ,
\end{equation}
where $\mathbf I$ is the nuclear spin, $\textbf{S}_{e(h)}$ is the steady-state polarization of the carriers induced by optical orientation. $l$ is the leakage factor characterizing the loss of nuclear polarization caused by relaxation processes other than the hyperfine coupling. Note, that a DNP develops only if $\textbf{S}_{e(h)}$ and $\mathbf{B}$ are not perpendicular to each other, as otherwise $\textbf{B}\cdot\textbf{S}_{e(h)}=0$. A detailed analysis of the DNP mechanism in the lead halide perovskites is given in Ref.~\onlinecite{kirstein2021}. The process is schematically illustrated in Figure~\ref{fig:DNP}b. The induced nuclear polarization $\braket{\textbf{I}}$ is collinear to $\mathbf{B}$. 

The polarized nuclear spins act back on the carriers via the Overhauser field $\mathbf{B}_{N}$, which is proportional to $\braket{\textbf{I}}$ as well as collinear to it and, thus, to $\mathbf{B}$. The Overhauser field adds to the external magnetic field, changing the frequency of the carrier spin precession in the total field $B\pm B_{N,e(h)}$, as $\omega_\textrm{L,e(h)}= |g_{e(h)}|\mu_\textrm{B}(B\pm B_{N,e(h)})/\hbar$. The magnitude of this change is a measure of the nuclear spin polarization or $\mathbf{B}_{N,e(h)}$. The direction of $\textbf{B}_{N,e(h)}=A_{e(h)} \langle \mathbf I\rangle/(g_{e(h)} \mu_{\rm B})$ is determined by the sign of the hyperfine coupling constant $A_{e(h)}$, the carrier $g$-factor, and the direction of $\braket{\textbf{I}}$. The last one is governed by the optically oriented $\mathbf{S}_{e(h)}$, see eq.~\eqref{eq:DNP}, and can be adjusted by varying the pump helicity. 

Figure~\ref{fig:DNP}a shows experimental results for the DNP in the MAPbI$_3$ crystal measured at $T=1.6$~K in a titled magnetic field of 100~mT. The two signals measured for opposite pump helicities differ considerably from each other in the period of the Larmor precession. Decomposition of these signals into the hole and electron components given below shows that for a $\sigma^+$ pump the hole Larmor frequency decreases, which means that for the holes the Overhauser field is opposite to the external magnetic field ($B_{N,h}<0$) which is in line with the upper right scheme in Figure~\ref{fig:DNP}b. For the electrons the Larmor precession frequency increases, i.e. for them the Overhauser field is collinear with $\mathbf{B}$ ($B_{N,e}>0$), see the upper left scheme in Figure~\ref{fig:DNP}b. We calculate the value of the Overhauser field by $|B_N|=|\hbar[\omega_{\rm L}(\sigma^+)-\omega_{\rm L}(\sigma^-)]/[2 g\mu_B]|$. For $\sigma^+$ pump $B_{N,h}=-10.8$~mT ($\omega_{{\rm L},h}(\sigma^-)=5.1$ rad ns$^{-1}$ and $\omega_{{\rm L},h}(\sigma^+)=3.9$~rad ns$^{-1}$) and $B_{N,e}=+1.7$~mT ($\omega_{{\rm L},e}(\sigma^-)=23.2$~rad ns$^{-1}$ and $\omega_{{\rm L},e}(\sigma^+)=24.0$~rad ns$^{-1}$). The $g$-factors for this field orientation ($\theta=60^\circ$, $\varphi=0^\circ$) measured in absence of a DNP under modulated pump helicity are $g_{h}=-0.68$ and $g_{e}=+2.78$. Note that irrespective to the pump helicity the sign of the Overhauser fields for electrons and holes is always different, due to their difference in $g$-factor sign.

We measured the dynamics of the nuclear spin system in respect of its repolarization when after establishing DNP the pump helicity is changed to the opposite polarization. The result is shown in Figure~\ref{fig:DNP}c. The KR intensity is detected at the delay time of 3.3~ns (compare with Figure~\ref{fig:DNP}c). The sample was illuminated by $\sigma^+$ pump during 30 minutes, then at time zero the pump polarization was switched for $\sigma^-$. The KR signal changes with a characteristic repolarization time of $T_{N}=16$ minutes, which represents an important parameter of the hyperfine interaction dynamics. It can be considered as a lower limit of the nuclear spin relaxation time $T_{1,N}$ under illumination. Typically, in semiconductors, the nuclear spin relaxation without illumination, which provides additional spin relaxation pathways, can be considerably longer.

\section{Spin relaxation of carriers at zero magnetic field} 

\begin{figure*}[!ht]
  \includegraphics[width=\textwidth]{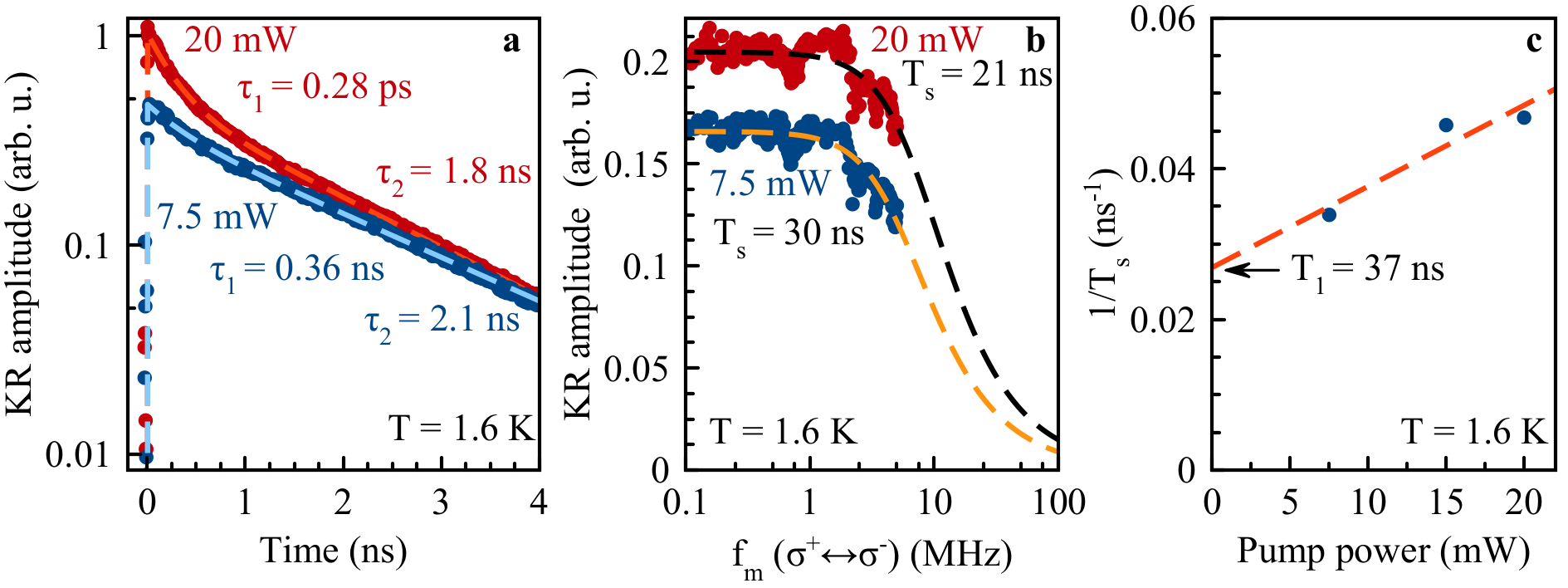}
  \caption{Longitudinal spin relaxation of carriers at zero magnetic field for $T=1.6$~K. (a) Time-resolved Kerr rotation measured for the pump powers of 7.5~mW (blue symbols) and 20~mW (red symbols). Dashed lines show double exponential fits with decay times given in the panel. (b) Spin inertia signals measured as the KR amplitude dependence on the helicity modulation frequency of the pump. The KR signal is detected at a fixed time delay of 1.5~ns for the pump powers of 7.5~mW (blue symbols) and 20~mW(red symbols). Dashed lines are fits after eq.~\eqref{eq:spininertia}. (c) Pump power dependence of the spin relaxation rate. Symbols give experimental data and line is the linear interpolation with eq. \eqref{eq:spininertia2} and $T_1=37$~ns. }
  \label{fig:spininertia}
\end{figure*}

In order to measure the carrier spin dynamics at zero magnetic field at long time scales we use the spin inertia technique~\cite{heisterkamp2015,smirnov2018}, which is based on TRKR. The spin inertia technique allows one to measure the longitudinal spin relaxation time $T_1$ by varying the modulation frequency of the pump helicity, $f_m$, and identifying the characteristic modulation frequency which exceeds the spin relaxation rate $1/T_1$, as described in \textbf{methods}.

In Figure~\ref{fig:spininertia}a the dynamics of the KR signal in zero magnetic field, $T=1.6$ and measured at the photon energy of 1.6402~eV is shown for two pump powers. In absence of a magnetic field the Larmor precession is turned off. The time evolution is fitted with a double exponential function providing decay times of 0.36~ns and 2.1~ns for $P=7.5$~mW as well as 0.28~ns and 1.8~ns for $P=20$~mW. The initial fast component is comparable with the $\tau_\textrm{X}=0.3$~ns of the exciton decay measured above by time-resolved differential reflectivity, see Figure~\ref{fig:exgfactor}d. The longer component of the spin signal evidences that it is contributed by long-living resident carriers. 

For evaluation of time scales exceeding the laser repetition frequency spin inertia curves were recorded, Figure~\ref{fig:spininertia}b. While the KR signal remains constant for low frequencies $f_m$ at a certain frequency the signal drops. The spin inertia curves were fitted using $A_{\rm KR} \propto \sqrt{\frac{1}{1+(2\pi f_m T_S)^2}}$, with $T_S$ the spin inertia cutoff time. 

We evaluate the time $T_1=37$~ns by extrapolation of the $T_S$ dependence on power to zero power, see the dashed line in Figure~\ref{fig:spininertia}c. Note that an assignment of this value to either the electrons or the holes is difficult, as spin beats caused by Larmor precession are absent in the Faraday geometry.

\section{Summary}

\begin{table}%
\caption{Measured and evaluated parameters for the MAPbI$_3$ single crystal. The spin dephasing time $T_2^*$ was measured at $B_{\rm V}=0.5~$T, $T=7$~K, and $P=0.25$~mW. Note that the longitudinal spin relaxation time $T_1$ cannot be unambiguously assigned to either the electron or the hole spin.}
\begin{tabular}{l|c|c}
    & hole & electron \\ \hline \hline 
$g_{\rm min}$ & $-0.28$& $+2.46$ \\
$g_{\rm max}$ & $-0.71$& $+2.98$\\
$T_2^*$ (ns) &2.7&0.4 \\
$T_1$ (ns) & $\le37$ & $\le37$ \\
$\Delta g$ &$0.003-0.008$& $0.03-0.04$ \\
$\Delta g/g$ ($\%$)&$0.9-1.2$& $1.2-1.4$\\
$B_{N}$ (mT) & $-10.8$ & $+1.7$
\end{tabular}
\label{tab:sum}
\end{table}

We have collected for convenience in Table~\ref{tab:sum} the key parameters characterizing the spin-dependent properties in the studied MAPbI$_3$ crystal. The measured $g$-factors for the electrons and holes are in good agreement with the recently established universal dependence of the charge carrier $g$-factors on the band gap in lead halide perovskites~\cite{kirstein2021nc}. In general, the experimental dependencies in the MAPbI$_3$ crystal measured with the approaches based on time-resolved Kerr rotation, including dynamic nuclear polarization and spin inertia, are in line with what was reported for CsPbBr$_3$ and FA$_{0.9}$Cs$_{0.1}$PbI$_{2.8}$Br$_{0.2}$ single crystals \cite{belykh2019,kirstein2021}. Also polycrystalline films, like MAPbI$_{3-x}$Cl$_x$~\cite{odenthal2017}, MAPbI$_3$~\cite{garcia-arellano2021} and CsPbBr$_3$~\cite{grigoryev2021}, despite of their larger inhomogeneity and spectral broadening, do not show drastic differences from single crystals. 

Comparing our results for the MAPbI$_3$ crystal with the data for a MAPbI$_3$ polycrystalline film~\cite{garcia-arellano2021}, one can see that the $g$-factor values in the film of $|g_{e}|=2.59-2.72$ and $|g_{h}|=0.24-0.35$ are close to those in the single crystal. While in the single crystal, the $g$-factor dispersion for both electrons and holes is about 1\%, in the film it reaches $\Delta g_{h}/g_{h} = 49\%$~\cite{garcia-arellano2021} for holes and $\Delta g_{e}/g_{e} = 0.6\%$~\cite{garcia-arellano2021} for electrons. 

The measured $T_1=37$~ns in MAPbI$_3$ is in the same range as the times reported for other lead halide perovskite single crystals, e.g., $T_1=53$~ns in CsPbBr$_3$~\cite{belykh2019} and 45~ns in FA$_{0.9}$Cs$_{0.1}$PbI$_{2.8}$Br$_{0.2}$~\cite{kirstein2021}. This evidences that the spin relaxation of charge carriers at a liquid helium temperature is controlled by their interaction with lead nuclear spins. Further experimental and theoretical arguments for that can be found in Ref.~\onlinecite{kirstein2021}. The observed efficient dynamic nuclear polarization via spin polarized carriers gives additional evidence for localization of the resident holes and electrons. 

\section{Methods}

\textbf{Samples.} Methylammonium (MA/CH$_3$NH$_3$) lead iodine (MAPbI$_3$) single crystals were low temperature solution grown in a reactive inverse temperature crystallization (RITC) process, Ref~\cite{hocker2021}. As compared to the pure inverse temperature crystallization, instead of pure $\gamma$-butyrolactone GBL precursor solvent an alcohol-GBL mixture was used. The mixed precursor solvent polarity is changed causing a lower solubility of MAPbI$_3$ and an optimization of the nucleation rate and centers, resulting in an earlier crystallization as compared to conventional growth temperatures. Black MAPbI$_3$ single crystals were grown at a temperature of $85^\circ$C instead of $110^\circ$C. At room temperature a tetragonal crystal phase with the lattice constants $a=8.93~$\r{A}, $c=12.5~$\r{A} was determined with XRD~\cite{hocker2021}.

\textbf{Optical measurements.} The sample was placed in a cryostat with temperatures variable from 1.6~K up to 300~K. For $T=1.6$~K, the sample is immersed in superfluid helium, while for the temperatures from 4.2~K to 300~K the sample is in cooling helium gas. The cryostat equipped with superconducting vector magnet was used. It has three orthogonal pairs of split coils to orient the magnetic field up to 3~T along any chosen direction. 

The magnetic field parallel to the light wave vector $\textbf{k} \parallel z$ is denoted as $\textbf{B}_{\rm F}$ (Faraday geometry) and the field perpendicular to $\textbf{k}$ as $\textbf{B}_{\rm V}$ (Voigt geometry). If not stated otherwise, $\textbf{B}_{\rm V}$ is chosen to be parallel to the $x$-axis, $\textbf{B}_{{\rm V},x}$. The angle $\theta$ is defined as the angle between $\textbf{B}_{\rm F}$ and $\textbf{B}_{\rm V}$, where $\theta=0^\circ$ corresponds to $\textbf{B}_{\rm F}$. The angle $\varphi$ describes the rotation in the Voigt/Voigt plane, which is perpendicular to $\textbf{k}$, with $\varphi=0^\circ$ corresponding to $\textbf{B}_{{\rm V},x}$ and $\varphi=90^\circ$ to $\textbf{B}_{{\rm V},y}$. A sketch of the experimental geometry and field orientations in shown in Figure~\ref{fig:aniso}a. 

\textbf{Photoluminescence and reflectivity}: The sample was excited for the photoluminescence measurements with a 405 nm (3.06~eV) continuous-wave (cw) diode laser with a low power of 1.2 mW and for reflectivity with an unpolarized white light lamp. The light coming from the sample (emission or reflection) was detected by a 0.5 m monochromator with a charge-coupled-device (CCD) camera. For polarization selective measurements directly after cryostat windows, before entering the fiber coupled entrance slit of the monochromator, a motorized circular retarder $\lambda/4$ together with a Glan Prism was placed. 

\textbf{Pump-probe time-resolved Kerr rotation (TRKR).} 
The coherent spin dynamics is measured by a degenerate pump-probe setup, where the pump and probe pulses have the same photon energy~\cite{yakovlevCh6}. A titanium-sapphire (Ti:Sa) laser generates 1.5-ps long pulses in the spectral range of $700-980$~nm ($1.265-1.771$~eV) with a spectral width of about 1~nm (about 1.5~meV) and a pulse repetition rate of 76~MHz (repetition period $T_\text{R}=13.2$~ns). The laser photon energy is tuned in the vicinity of the exciton resonance at 1.6365~eV, where the TRKR signal is strongest. The laser emission is split into two beams (the pump and the probe). The probe pulses are delayed relative to the pump pulses by a mechanical delay line. Both the pump and probe beams are modulated using a photo-elastic modulators (PEM). The probe beam is linearly polarized and amplitude modulated at a frequency of 84~kHz, while the pump beam is either helicity modulated at 50~kHz between the $\sigma^+/\sigma^-$ circular polarizations, or has constant helicity, either $\sigma^+$ or $\sigma^-$, and is amplitude modulated. 

The polarization of the reflected probe beam is analyzed, using a lock-in technique, in respect of full set of Stokes polarimetry, its Kerr rotation (KR) or its Kerr ellipticity (KE). For the KE [KR], the circular [linear] polarization was analyzed ($\lambda/4$ [$\lambda/2$] retarder together with Wollaston prism)~\cite{glazov2010}. In a finite magnetic field, the Kerr amplitude oscillates in time reflecting the Larmor spin precession of the carriers and decays as the time delay between pump and probe increases. When both electrons and holes contribute to the KR signal, as is the case for the studied perovskite crystal, the signal can be described with a superposition of two decaying oscillatory functions: $A_{\rm KR}(t) = S_{e} \cos (\omega_{{\rm L},e} t) \exp(-t/T^*_{2,e}) + S_{h} \cos (\omega_{{\rm L},h} t) \exp(-t/T^*_{2,h})$. The envelope of the signal gives the spin dephasing times $T_2^*$. For standard TRKR measurements, the external magnetic field is applied in the Voigt geometry perpendicular to the light wave vector $\textbf{k}$. The angle between them was varied for measuring the $g$-factor anisotropy. The $g$-factors are evaluated from $\omega_{{\rm L},e(h)}$ using the relation $|g_{e(h)}|= \hbar \omega_{{\rm L},e(h)} / (\mu_{\rm B} B)$. 

\textbf{Time-resolved differential reflectivity (TRDR).} The exciton lifetime was determined by TRDR. In these measurements a pulsed laser beam (same Ti:Sa laser as in TRKR) was tuned to the reflection minimum at 1.638~eV. The laser beam was split into three beams: pump, probe and reference. The probe is directed after reflection from the sample to a balanced photodetector, while the reference beam is guided directly to another channel of the detector. The reference and probe intensities were balanced on the photodetector to obtain zero signal when the pump was blocked. Applying the pump pulse to the sample droves the balanced beam out of equilibrium by the injected exciton population. The exciton population dynamics was measured by varying the time delay between the pump and probe pulses. All beams were linearly polarized.

\textbf{Spin inertia.} The spin polarization process depends on the relaxation time of the spins $T_1$. In the experiment a helicity modulated excitation is chosen. For a period $1/(2f_m)$ the spin polarization is build up with $\sigma^+$, followed by a reversed period with $\sigma^-$ excitation. The obtained Kerr signal $A_{\rm KR}$ is the integral over those periods under assumption of $T_1 \ll 1/f_m$. If the modulation frequency is increased in such manner that this approach is not longer fulfilled one can show~\cite{heisterkamp2015,smirnov2018}:
\begin{equation}
\label{eq:spininertia}
A_{\rm KR} \propto \sqrt{\frac{1}{1+(2\pi f_m T_S)^2}}.
\end{equation}
Further one need to take into account a faster reaching of the equilibrium spin polarization with higher pumping power $P$ and the spin generation coefficient $G$, 
\begin{equation}
\label{eq:spininertia2}
\frac{1}{T_S} = \frac{1}{T_1} + GP.
\end{equation}
The power corrected spin inertia cutoff time $T_S$ gives a good estimate of the longitudinal spin relaxation time $T_1$.

\textbf{AUTHOR INFORMATION}

Corresponding Authors:

Email: erik.kirstein@tu-dortmund.de\\ 
Email: dmitri.yakovlev@tu-dortmund.de\\

\textbf{ORCID}\\
Erik Kirstein:        0000-0002-2549-2115 \\
Dmitri R. Yakovlev:   0000-0001-7349-2745 \\
Evgeny A. Zhukov:     0000-0003-0695-0093 \\
Julian H\"ocker:   0000-0002-8699-3431\\
Vladimir Dyakonov:    0000-0001-8725-9573\\
Manfred Bayer:        0000-0002-0893-5949\\

\textbf{Notes}

The authors declare no competing financial interests.

\section{Acknowledgments}

The authors are thankful to M. M. Glazov, N. E. Kopteva, A. Greilich and I. A. Akimov for fruitful discussions.
We acknowledge the financial support by the Deutsche Forschungsgemeinschaft in the frame of the Priority Programme SPP 2196 (Project YA 65/26-1 and DY 18/15-1) and the International Collaboration Research Center TRR160 (project A1). J.H. and V.D. acknowledge financial support from the Deutsche Forschungsgemeinschaft through the W\"urzburg-Dresden Cluster of Excellence on Complexity and Topology in Quantum Matter-ct.qmat (EXC 2147, project-id 39085490).



\end{document}